\documentclass[12pt]{article}
\usepackage{amsfonts,amsmath}
\advance\textheight by \topskip
\begin{document}
\vskip 5mm
\begin{center}
{\Huge
An algebraic method of obtaining of symplectic
coordinates in a rigid body dynamics}
\vskip 5mm
     {\large Alexander Pavlov}
\vskip 5mm
Department of Theoretical Mechanics\\
Agricultural Academy\\
11 Studencheskaya Street\\
426069, Izhevsk, RUSSIA\\
\end{center}
\vskip 5mm
\abstract{}
An algebraic procedure of getting of canonical variables in a rigid body
dynamics is presented. The method is based on using
a structure of an algebra of Lie---Poisson
brackets with which a Hamiltonian dynamics is set.
In a particular case of a problem of a top in a
homogeneous gravitation field the method leads to well--known
Andoyer---Deprit variables.
Earlier, the method of getting of them was based on
geometric approach (Arhangelski, 1977).
\section*{Introduction}

Poisson brackets play a key role in Hamiltonian mechanics. According to
the theorem of Darboux (Olver, 1986), degenerated Poisson manifolds are
stratified to
symplectic manifolds (leaves). A reduction of a Hamiltonian system on common
level of all Casimir functions leads to usual Hamiltonian mechanics.
The reduction is especially algebraic problem and can be made without
referring to a concrete physical problem. In the present paper an algebraic
method of finding of symplectic coordinates of Lie---Poisson brackets in some
important for physical application cases is demonstrated.

\section{A symplectic basis of the Lie---Poisson brackets with an algebra
$e(3)$}
\setcounter{section}{1}
\setcounter{equation}{0}
The equations of motion of a rigid body with a fixed point in a homogeneous
gravitation field are traditionally written in variables
${\bf M},{\boldsymbol\gamma},$
where ${\bf M}$ is a vector of an angular momentum of the body and
${\boldsymbol\gamma}$ is unit vector parallel to the vector of
gravitational field strength.
The coordinate system is toughly connected with the body.

The equations of motion are Hamiltonian with a degenerated Poisson brackets
(Olver, 1986)
\begin{equation} 
{\{M_i,M_j\}=-\epsilon_{ijk}M_k,\quad
\{\gamma_i,\gamma_j\}=0,\quad
\{M_i,\gamma_j\}=-\epsilon_{ijk}\gamma_k}
\end{equation}

The algebra (1.1) is a semi--direct sum of an algebra of rotation $so(3)$
and an algebra of translation R$^3:$
$e(3)=so(3)\oplus_{s}$R$^3.$
Under reduction of the algebra (1.1)
to  Casimir functions $H=({\bf M},{\boldsymbol\gamma})$ and
$({\boldsymbol\gamma},{\boldsymbol\gamma})=1,$ the Poisson brackets (1.1)
become  non--degenerated
and the equations of motion can be written in canonical form.
In the capacity of canonical coordinates, most convenient for qualitative
analysis, the canonical variables of Andoyer---Deprit $(l,L;g,G;h,H)$  are used.

For describing of mechanical sense of these variables we denote by
$OXYZ$ a stationary trihedron with the origin in the point of fixation of the
body, $Oxyz$ is connected with the body the frame of coordinates,
$\Sigma$ is a plane passing through the point of fixation perpendicular
to the vector of angular momentum ${\bf M}.$ At last, in the admitted
notations

$L$ is a projection of the angular momentum to the moving axis $Oz$;

$G$ is a value of the angular momentum;

$H$ is a projection of the angular momentum to the fixed axis $OZ$;

$l$ is an angle between the axis $Ox$ and a line of intersection of
$\Sigma$ with $Oxy$;

$g$ is an angle between the line of intersection of $\Sigma$ with planes
$Oxy$ and $OXY$;

$h$ is an angle between the axis $OX$ and the line of intersection of
$\Sigma$ with the plane $OXY.$

The essence of the method consists in extracting from (1.1) of the subalgebra
$so(3)$ and getting of canonical variables there.
The canonical variables $l,$ $L$
\begin{equation}  
M_1=\sqrt{G^2-L^2}\sin{l}, \quad M_2= \sqrt{G^2-L^2}\cos{l},\quad M_3=L,
\end{equation}
are cylindrical coordinates on two--dimensional
concentric spheres (orbits of the $so(3))$ (Olver, 1986).
The angular momentum $G=\sqrt{M_1^2+M_2^2+M_3^2}$
is the Casimir function of the considered algebra.
The Poisson brackets between
$l, L, G, \gamma_1, \gamma_2, \gamma_3$ are following:
\begin{equation} 
{\{l,L\}=1,\quad \{G,L\}=\{G,l\}=0,\quad\{\gamma_i,\gamma_j\}=0,}
\end{equation}
\vskip 3mm
\begin{equation} 
{\{L,\gamma_1\}=-\gamma_2,\quad \{L,\gamma_2\}=\gamma_1,\quad
\{L,\gamma_3\}=0,}
\end{equation}
\vskip 3mm
$${\{l,\gamma_1\}=-\frac{\sin l\gamma_3}{\sqrt{G^2-L^2}},\quad
\{l,\gamma_2\}=-\frac{\cos l\gamma_3}{\sqrt{G^2-L^2}}},$$
\begin{equation} 
\{l,\gamma_3\}=\frac{H-L\gamma_3}{G^2-L^2},
\end{equation}
\vskip 3mm
$$\{G,\gamma_1\}=\frac{1}{G}(\sqrt{G^2-L^2}\cos l\gamma_3-L\gamma_2),$$
\begin{equation} 
\{G,\gamma_2\}=\frac{1}{G}(L\gamma_1-\sqrt{G^2-L^2}\sin l\gamma_3),
\end{equation}
$$\{G,\gamma_3\}=\frac{1}{G}\sqrt{G^2-L^2}(\sin l\gamma_2-\cos l\gamma_1),$$
where $H$ is a projection of the angular momentum to the fixed axis.
In addition, $H=({\bf M}, {\boldsymbol\gamma})$ is the Casimir function of the
algebra $e(3).$

Now we solve step by step the systems of partial differential equations
(1.2)--(1.6), assuming $\gamma_i$ as functions of $(l,L;g,G;H).$
As a result, we get the required relations with the Andoyer---Deprit variables:
\begin{equation} 
\gamma_1=\left(\frac{H}{G}\sqrt{1-{\left(\frac{L}{G}\right)}^2}+
\frac{L}{G}\sqrt{1-{\left(\frac{H}{G}\right)}^2}\cos g\right)\sin l+
\sqrt{1-{\left(\frac{H}{G}\right)}^2}\sin g \cos l,
\end{equation}
$$\gamma_2=\left(\frac{H}{G}\sqrt{1-{\left(\frac{L}{G}\right)}^2}+
\frac{L}{G}\sqrt{1-{\left(\frac{H}{G}\right)}^2}\cos g\right)\cos l-
\sqrt{1-{\left(\frac{H}{G}\right)}^2}\sin g \sin l,$$
$$\gamma_3=\left(\frac{H}{G}\right)\left(\frac{L}{G}\right)-
\sqrt{1-{\left(\frac{L}{G}\right)}^2}
\sqrt{1-{\left(\frac{H}{G}\right)}^2}\cos g.$$

The canonical variables of the four--dimensional symplectic leaf of the
Poisson brackets of the algebra
$e(3)$ are $(l,L;g,G).$ They are enumerated by the variable $H.$

\section{A symplectic basis of the Lie---Poisson brackets with an algebra
$l(7)$}
\setcounter{section}{2}
\setcounter{equation}{0}
The very convenient variables for description of the rigid body dynamics
instead of the angles $({\boldsymbol\alpha}, {\boldsymbol\beta},
{\boldsymbol\gamma})$ that form
the orthogonal matrix of rotations are quaternions (Koshlyakov, 1985). The real
variables $\lambda_0, \lambda_1, \lambda_2, \lambda_3$
normalized by the condition
\begin{equation} 
\lambda_0^2+\lambda_1^2+\lambda_2^2+\lambda_3^2=1,
\end{equation}
and connected with the vector ${\boldsymbol\gamma}$ by the formulae (Koshlyakov, 1985):
$$\gamma_1=2(\lambda_1\lambda_3-\lambda_0\lambda_2),$$
\begin{equation} 
\gamma_2=2(\lambda_0\lambda_1+\lambda_2\lambda_3),
\end{equation}
$$\gamma_3=\lambda_0^2-\lambda_1^2-\lambda_2^2+\lambda_3^2,$$
are named the Hamilton---Rodrigues parameters.

A point M({\bf r}) under some rotation around a vector
${\bf  e}(\alpha',\beta',\gamma')$
is displaced to ${M}'({\bf r}')$ on an angle $\chi.$ The vector
${\boldsymbol\theta}=2 tan(\chi/2){\bf e}$ is a vector of finite rotation.
Instead of projections $\theta_1, \theta_2, \theta_3$ one can introduce
variables $\lambda_k,$ by $\lambda_k=1/2\lambda_0\theta_k,$ $k=1,2,3$
are subjected to the condition (2.1). So we get
\begin{equation} 
\lambda_0=\cos{\chi/2},\quad\lambda_1=\cos\alpha'\sin{\chi/2},\quad
\lambda_2=\cos\beta'\sin{\chi/2},\quad\lambda_3=\cos\gamma'\sin{\chi/2}.
\end{equation}

Now, it is not difficult to obtain the algebra of Poisson brackets of
variables
${\bf M},\lambda_0,\lambda_1,\lambda_2,\lambda_3.$ The Lagrangian of a
free rigid body with a diagonal tensor of inertion
${\bf I}=diag(A,B,C)$ is
\begin{equation} 
L=\frac{1}{2}({\bf I}{\boldsymbol\omega},{\boldsymbol\omega}),
\end{equation}
where ${\boldsymbol\omega}$ is a vector of angular velocity.
In capacity of quasimomenta we take components of the angular momenta
\begin{equation} 
{\bf M}={\partial L}/{\partial {\boldsymbol\omega}}.
\end{equation}
The Hamiltonian ${\cal H}$ is defined by the Legandre transformation
\begin{equation} 
{\cal H}=\left({\boldsymbol\omega},
\frac{\partial L}{\partial{\boldsymbol\omega}}\right)-L\mid_
{{\boldsymbol\omega}\to {\bf M}}.
\end{equation}

Using (2.5), (2.6) and kinematic Euler relations (Koshlyakov, 1985):
$$\omega_1=\dot\psi\sin\theta\sin\varphi+\dot\theta\cos\varphi,$$
\begin{equation} 
\omega_2=\dot\psi\sin\theta\cos\varphi-\dot\theta\sin\varphi,
\end{equation}
$$\omega_3=\dot\psi\cos\theta+\dot\varphi,$$
where $\theta,\varphi,\psi$ are the Euler angles,
we obtain the formulae:
$$M_1=\frac{\sin\varphi}{\sin\theta}(p_\psi-p_{\varphi} \cos\theta )+
p_\theta \cos{\varphi} ,$$
\begin{equation} 
M_2=\frac{\cos\varphi}{\sin\theta}(p_\psi-p_{\varphi} \cos\theta )-
p_\theta \sin\varphi ,
\end{equation}
$$M_3=p_\varphi.$$
Taking under consideration canonical brackets between generalized
coordinates~---Euler angles and corresponding to them canonical momenta
$p_\theta, p_\varphi, p_\psi,$
it is possible to obtain the algebra
\begin{equation} 
{\{M_{i},M_{j}\}=-\epsilon_{ijk}M_{k},\qquad
\{M_{i},\lambda_{0}\}=\frac{1}{2}\lambda_{i},}
\end{equation}
$${\{M_{i},\lambda_{j}\}=-\frac{1}{2}(\epsilon_{ijk}\lambda_{k}+\delta_{ij}
\lambda_{0}),
\qquad\{\lambda_{\mu},\lambda_{\nu}\}=0.}$$

Let us obtain by the developed method canonical coordinates of the
six--dimensional symplectic leaf of the Poisson brackets with a
seven--dimensional Lie algebra
$l(7)=so(3)\oplus_s $R$^4.$
Also, as in the first section, we solve step by step the following systems
of partial differential equations corresponding to the algebras.
\begin{equation} 
{1.\quad \{L,\lambda_0\}=\frac{1}{2}\lambda_3,\quad
\{L,\lambda_1\}=-\frac{1}{2}\lambda_2,}
\end{equation}
$${\{L,\lambda_2\}=\frac{1}{2}\lambda_1,\quad
\{L,\lambda_3\}=-\frac{1}{2}\lambda_0;}$$
\vskip 3mm
\begin{equation}  
2.\quad \{H,\lambda_0\}=\frac{1}{2}\lambda_3,\quad
\{H,\lambda_1\}=\frac{1}{2}\lambda_2,
\end{equation}
$$\{H,\lambda_2\}=-\frac{1}{2}\lambda_1,\quad
\{H,\lambda_3\}=-\frac{1}{2}\lambda_0;$$
\vskip 3mm
\begin{equation} 
3.\quad \{l,\lambda_0\}=\frac{\lambda_1 \cos l-\lambda_2 \sin l}
{2\sqrt{G^2-L^2}},\quad \{l,\lambda_1\}=
-\frac{\lambda_3 \cos l+\lambda_0 \sin l}{2\sqrt{G^2-L^2}},
\end{equation}
$$\{l,\lambda_2\}=\frac{\lambda_0 \sin l-\lambda_3 \cos l}
{2\sqrt{G^2-L^2}},\quad \{l,\lambda_3\}=
\frac{\lambda_1 \sin l+\lambda_2 \cos l}
{2\sqrt{G^2-L^2}};$$
\vskip 3mm
$$4.\quad \{G,\lambda_0\}=\frac{\sqrt{G^2-L^2}}{2G}
(\lambda_1 \sin l+\lambda_2 \cos l)+
\frac{L}{2G}\lambda_3,$$
\begin{equation} 
\{G,\lambda_1\}=\frac{\sqrt{G^2-L^2}}{2G}
(-\lambda_0 \sin l+\lambda_3 \cos l)-
\frac{L}{2G}\lambda_2,
\end{equation}
$$\{G,\lambda_2\}=-\frac{\sqrt{G^2-L^2}}{2G}
(\lambda_0 \cos l+\lambda_3 \sin l)+
\frac{L}{2G}\lambda_1,$$
$$\{G,\lambda_3\}=\frac{\sqrt{G^2-L^2}}{2G}
(\lambda_2 \sin l-\lambda_1 \cos l)-
\frac{L}{2G}\lambda_0.$$

Using the condition of the norm (2.1) and relations (2.2)
we get the following solutions:

$$\lambda_0=\frac{1}{\sqrt{2}}
(\sin (g/2) \sin (y_{+}) \cos (x_{-})+
\sin (g/2) \cos (y_{+}) \cos (x_{-})+$$
$$\cos (g/2) \sin (y_{+}) \sin (x_{+})-
\cos (g/2) \cos (y_{+}) \sin (x_{+})),$$
$$\lambda_1=\frac{1}{\sqrt{2}}
(\sin (g/2) \cos (y_{-}) \sin (x_{-})-
\sin (g/2) \sin (y_{-}) \sin (x_{-})-$$
$$\cos (g/2) \sin (y_{-}) \cos (x_{+})-
\cos (g/2) \cos (y_{-}) \cos (x_{+})),$$
\begin{equation}  
\lambda_2=\frac{1}{\sqrt{2}}
(-\sin (g/2) \sin (y_{-}) \sin (x_{-})-
\sin (g/2) \cos (y_{-}) \sin (x_{-})+
\end{equation}
$$\cos (g/2) \sin (y_{-}) \cos (x_{+})-
\cos (g/2) \cos (y_{-}) \cos (x_{+})),$$
$$\lambda_3=\frac{1}{\sqrt{2}}
(\sin (g/2) \sin (y_{+}) \cos (x_{-})-
\sin (g/2) \cos (y_{+}) \cos (x_{-})-$$
$$\cos (g/2) \sin (y_{+}) \sin (x_{+})-
\cos (g/2) \cos (y_{+}) \sin (x_{+})).$$

In the formulae (2.14) there are introduced the angles
$\zeta, \tau:$ 
$${\zeta= \arcsin H/G,\quad \tau= \arcsin L/G}$$
and the combinations:
$${x_{+}\equiv\frac{1}{2}(\zeta+\tau ),\quad
x_{-}\equiv\frac{1}{2}(\zeta-\tau ),\quad
y_{+}\equiv\frac{1}{2}(l+h),\quad y_{-}\equiv\frac{1}{2}(l-h ).}$$

The obtained formulae (2.14) can be used for applications of methods of the
theory of perturbation to the problem of the rigid body rotations
in superposition of some potential strength fields.
\newpage
\section*{References}
Arhangelski Yu.A. (1977). {\it Analytical dynamics of a rigid body},
Nauka, Moscow [in Russian].\\
Koshlyakov V.N. (1985). {\it Problems of dynamics of rigid body
and applied theory of gyroscopes: Analytical methods},
Nauka, Moscow [in Russian].\\
Olver P. (1986). {\it Applications
of Lie groups to differential equations}, Springer--Verlag, New York.
\end{document}